\documentclass[useAMS]{mn2e} 

\usepackage{amssymb,amsmath,psfig,times,graphicx,longtable}
\def\gsim{ \lower .75ex \hbox{$\sim$} \llap{\raise .27ex \hbox{$>$}} }
\def\lsim{ \lower .75ex\hbox{$\sim$} \llap{\raise .27ex \hbox{$<$}} }

\def\sc{Schwarzschild}
\def\beq{\begin{equation}}
\def\eeq{\end{equation}}

\def\fe{{\it Fermi}}
\def\sw{{\it Swift}}

\title[AT20G-1FGL]
{Correlation of \fe--LAT sources with the AT20GHz radio survey}  
\author[G. Ghirlanda, G. Ghisellini, F. Tavecchio, L. Foschini ]
{G. Ghirlanda\thanks{Email:
giancarlo.ghirlanda@brera.inaf.it}, G. Ghisellini, F. Tavecchio, L. Foschini \\
INAF -- Osservatorio Astronomico di Brera, Via Bianchi 46, I--23807 Merate, Italy\\
}
\begin{document}  

\maketitle

\begin{abstract}
We cross correlate the \fe\ 11 months survey catalogue (1FGL) 
with the 20 GHz Australia Telescope Compact Array radio survey catalogue  (AT20G)  
composed by 5890 sources at declination $<$0$^\circ$. 
Among the 738 \fe\ sources distributed in the southern sky we find 
230 highly probable candidate counterparts in the AT20G survey. 
Of these, 222 are already classified as blazars 
(176 of known type and 46 of unknown optical class) in the \fe\ 
1--year LAT AGN Catalogue (1LAC) and 8 are new associations.  
By studying the $\gamma$--ray and radio properties of these 
associations we find a strong correlation between the $\gamma$--ray 
flux (above 100 MeV) and the 20 GHz flux density. 
This correlation is more than 3$\sigma$ statistically significant both 
for the population of BL Lacs and of FSRQ considered separately. 
We also find that the radio counterparts associated to the \fe\ sources 
have on average flat radio spectra between 5 and 20 GHz and that \fe\ 
$\gamma$--ray sources are not preferentially associated with 
``ultra inverted spectrum" radio sources. 
For 2 of the 8 new associations we build the broad band spectral energy 
distribution combining \fe, \sw\ and radio observations. 
One of these two sources is identified with the high redshift
FSRQ Swift J1656.3--3302 ($z=2.4$)
and we classify the other source as a candidate new FSRQ. 
We also study the brightest radio source of 
the 46 associations without an optical classification and classify it 
as a new BL Lac candidate ``twin" of the prototypical BL Lac OJ 287 if 
its redshift is somewhat larger, $z\sim$0.4. 
\end{abstract}

\section{Introduction}

The Large Area Telescope (LAT) on board the \fe\ satellite (Atwood et al. 2009) detected 1451 
sources in the $\gamma$--ray band above 100 MeV with a significance $>$4.5$\sigma$ 
during its first 11 months survey (Abdo et al. 2010, A10 hereafter). 
This is the 1FGL 
catalogue\footnote{http://fermi.gsfc.nasa.gov/ssc/data/access/lat/1yr\_catalog/} 
and contains both galactic and extragalactic $\gamma$--ray sources. 
In a recent paper Abdo et al. (2010a, A10a hereafter) classified 831
out of 1451 1FGL sources as blazars. 
These form the 1LAC 
catalogue\footnote{http://heasarc.gsfc.nasa.gov/W3Browse/fermi/fermilac.html}. 
Considering the 796 sources in the 1LAC associated with a single counterpart, 
there are 37 sources generically classified as AGN,  314 BL Lac and 285 Flat Spectrum Radio Quasars (FSRQ), 
classified according to their optical spectrum. 
The remaining 160 sources are candidate blazars but of ``unknown" class 
because of lacking an optical spectrum or, if available, because of its 
poor quality for the optical classification. 
However, for these sources a counterpart could still be found in one of 
the radio catalogues adopted for the source identification by A10a 
(CRATES -- Haeley et al. 2007; CGRaBS -- Healey et al. 2008; the BZCat -- 
Massaro et al. 2009). 
Thirty five sources in the 1LAC sample have more than one associated counterpart. 

The 1FGL sample is equally populated in the northern (713 sources) 
and southern sky (738 sources). 
However, only 50\% of the 1FGL sources in the south sky are classified as 
blazars in the 1LAC catalogue with respect to $\sim$70\% of the 1FGL northern 
sources which are classified as blazars in the 1LAC catalogue. 

The recently published Australia Telescope 20 GHz Survey (AT20G, 
Murphy at al. 2010) represents the largest catalogue of high frequency 
radio sources detected with the Australia Telescope Compact Array (ATCA) 
in a survey conducted from 2004 to 2008 and covering the whole sky south 
of declination 0$^{\circ}$. 
It contains  5890 radio sources with flux at 20 GHz exceeding 40 mJy and 
it is complete at a level of 91\% above 100 mJy beam$^{-1}$. 
For several sources the fluxes at 8 and 5 GHz are also measured.  
The AT20G survey, being conducted at relatively high radio frequencies 
(20 GHz), most likely detects the radio emission from the compact cores of AGN. 
Several sources in this survey have flat or inverted radio spectra between 
5 and 20 GHz.

% It is worth to cross 
We correlate the AT20G radio catalogue with the 1FGL 
in search for possible counterparts. 
This 
% correlation 
allows us to study the radio versus $\gamma$--ray 
properties of these associations. 
A possible correlation between the radio and the $\gamma$--ray emission 
in radio loud AGN can shed light on the physical link between the emission 
processes in these two energy bands. 
%
% Indeed, while synchrotron emission 
% by relativistic electrons is thought to dominate in the radio band, the 
% high energy $\gamma$--ray emission can be dominated by Inverse Compton 
% scattering by relativistic electrons on seed photons of different origin 
% (e.g. Ghisellini \& Tavecchio 2008). 
%
The $\nu F_{\nu}$ spectra of both 
FSRQ and BL Lac objects show two peaks which are widely interpreted as 
due to the synchrotron and inverse Compton emission, respectively.  
The Spectral Energy Distribution (SED) of blazars form a sequence 
(so--called ``blazar sequence", e.g. Fossati et al. 1998) with the 
most powerful sources (FSRQs) having the synchrotron and Inverse Compton 
peaks in the IR and MeV energy range respectively, while the less powerful 
BL Lacs have the synchrotron and Inverse Compton peak in the 
optical/UV (or even X--ray) and GeV--TeV energy range respectively. 
The blazar sequence was built by dividing blazars into bins of radio luminosity,
thought to be a proxy for the bolometric one, and establishes a 
link between the radio and the $\gamma$--ray emission.
% The blazar sequence establishes a link between the radio and the $\gamma$--ray 
% emission in these sources. 

Before \fe, no conclusive claim could be made on the existence of a 
possible correlation between radio and $\gamma$--ray properties for 
EGRET sources (e.g. Mucke et al. 1997). 
Taylor et al. (2007), by studying the radio properties of EGRET detected 
blazars with the VLBA Imaging and Polarimetry Survey 
(VIPS -- extending down to 85 mJy) found that at low radio fluxes the radio 
flux density does not directly correlate with the $\gamma$--ray flux. 
Further investigation of this correlation was possible with the source 
list of the \fe\ first three months survey (LBAS, Abdo et al. 2009; 
Giroletti et al. 2010): while a correlation between the radio flux at 8.4 GHz 
and the peak flux (above 100 MeV) appears when considering BL Lacs and 
FSRQ together, the statistical evidence is only marginal for FSRQs
(chance correlation probability of 8\%), being more robust only for BL Lacs (0.5\%). 
Such correlation analysis should consider the different redshift range 
spanned by these two populations (e.g. Mucke et al. 1997). 
Abdo et al. (2009) found a  separation in the $\gamma$--ray spectral index 
versus radio luminosity among FSRQs, BL Lacs and misaligned AGN which suggests 
that more luminous radio sources have softer $\gamma$--ray spectra.  
Recent studies of the radio--$\gamma$ flux correlation in the LBAS sources
% \fe\ first three months source sample 
(Kovalev et al. 2009; 2009a) was conducted 
using 
% by comparing 
the MOJAVE sample of extragalactic sources (with a flux limit of 1.5 Jy at 15 GHz).
% and the corresponding blazars in the LBAS. 
Kovalev et al. (2009) find that the parsec--scale radio emission and the 
$\gamma$--ray flux are strongly related in bright $\gamma$--ray objects,
% detected by \fe\ 
suggesting that \fe\ selects the brightest objects from 
a flux--density limited sample of radio--loud sources. 

In this paper we search for the possible counterparts of the \fe\ 
1FGL sample in the AT20G survey (\S 1). 
By intersecting the association results with the 1LAC sample we study 
the radio vs $\gamma$--ray properties of these associations (\S 2). 
The 1FGL--AT20G correlation also reveals 8 new associations which are 
unclassified sources in the 1FGL sample. 
For two of these there are \sw\ observations which, combined with the radio and 
\fe\ data, allow to build the broad band SED. One of these two sources is classified here as a new candidate
 FSRQ (\S 3). We also present, as a first result of an on--going study of the properties 
of the 46 unknown sources of the 1LAC sample, the SED of the brightest 
radio source of the UIS class that we associate with one 1FGL source. 
We classify this object as a BL Lac 
due to the similarity of its SED to that of OJ 287 (\S 4). We discuss our findings in \S 5.

\section{\fe--LAT 1FGL / AT20G cross correlation}

One method used to find \fe\ counterparts (Abdo et al. 2009, 2010, 2010a) 
relies on positional coincidence (Sutherland \& Saunders 1992) and it 
is based on Bayes statistics. To each association between a \fe\ source 
and a candidate counterpart it is assigned a posterior probability based 
on the counterparts' space density around the \fe\ source. 
Only those associations with a  posterior probability larger than 
a chosen value are considered as likely associations. 
Considering only high latitude sources ($|b|>10^{\circ}$) the 
1LAC sample (Abdo et al. 2010) contains 709 AGN counterparts 
(i.e. also multiple counterparts within the \fe\ source error ellipse) 
associated to 671 \fe\ 1FGL sources at high galactic latitudes with 
posterior probabilities $>50$\%. Among these, 663 have posterior 
probabilities $>$80\% and 599 are defined as "clean" associations if 
(i) they have one single associations; 
(ii) the posterior probability is larger than 80\% and 
(iii) are not flagged as problematic (e.g. marginal detection or data problems) 
in the 1FGL sample. 

The method adopted by A10a to build the 1LAC sample is 
implemented in the \texttt{gtsrcid} tool distributed with 
the \texttt{LAT ScienceTools-v9r15p2} software.  
The association method searches for single radio sources 
falling within the confidence ellipse of the \fe\ sources. 
We applied this method in cross correlating the AT20G survey 
catalogue with the 1FGL one. Similarly to the 1LAC sample we accept 
only single associations (i.e. only one radio counterpart falls 
within the the \fe\ source error ellipse) and with a posterior probability $>$80\%. 

\subsection{Prior probability of the AT20G survey}

To compute the posterior probability we have to assign 
the prior probability to the AT20G sample. 
This  represents the confidence that the AT20G catalogue contains 
the real counterparts of the \fe\ sources. 
We compute the prior 
probability of the AT20G catalogue following the procedure described 
in Appendix C of A10. 
The prior probability represents the probability 
of finding a number of false identification from the correlation of the 
1FGL and AT20G real catalogues which is equal to the average number of 
associations found by the cross correlation of the AT20G sample 
with randomly generated catalogues of \fe\ sources. 
To this aim we created a set of 100 \fe\ fake catalogues by displacing 
the sources between 2 and 10 degrees from their real positions. 
For sources near the galactic plane a smaller displacement in 
galactic latitude was adopted (Eq. C3 of Appendix 2 of A10).

\begin{figure}
\hskip -1.4 cm
\psfig{figure=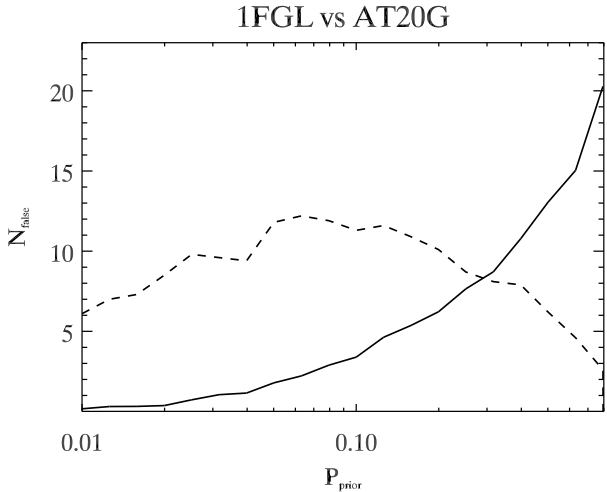,width=9.8cm,height=7cm}
\vskip -0.5 cm
\caption{
Number of false associations as a function of the prior probability 
assumed in cross correlating the AT20G survey with the \fe\ 11 
months source catalogue (1FGL) (dashed line). 
The solid line represents the average number of associations 
found by a cross correlation of the AT20G with 100 random 
\fe\ catalogues (for each value of the prior probability). 
The intersection between the two curves represents the 
prior probability used in the cross correlation of the AT20G survey and the 
1FGL catalogue (see A10).}
\label{fg1}
\end{figure}

We then computed the number of false associations between the AT20G 
and 1FGL as a function of the prior probability assigned to the AT20G 
catalogue (dashed line in Fig. \ref{fg1}) and the average number of 
associations between the AT20G and 100 1FGL random catalogues for each 
value of the prior probability (solid line in Fig. \ref{fg1}). 
The intersection between the two curves in Fig. \ref{fg1} gives 
the value of the prior probability adopted for the AT20G sample.

%----------------------------------------------------
%----------------------------------------------------

\subsection{Cross correlation results}

From the cross correlation of the AT20G and the 1FGL catalogue we 
found 230 associations for the 738 1FGL sources distributed in the 
southern sky, i.e in the same sky portion covered by the AT20G survey. 
Among these associations  222  sources are already classified as blazars 
in the 1LAC sample: 10 are generically classified as AGN, 54 are BL Lac, 112 are FSRQ 
and 46 are of unknown type. 
There are 8 remaining sources which are newly  found associations, i.e. 
they are not present in the 1LAC sample. 
Tab. 3 lists the properties of the 222 1FGL--AT20G associations 
which are classified as blazars in the 1LAC sample. 
In Tab. 2 we report the 8 new associations.  
For each association we give its posterior probability, the separation 
between the \fe\ 1FGL source centroid and the position of the AT20G 
associated source, the 1FGL and AT20G names,
the name of the AGN that was associated to the \fe\ 1FGL source in the 
1LAC sample and its optical classification (AGN, BLL, FSRQ or UNKNOWN) 
and if the 1LAC association was classified as ``clean" or not. We note that in the 
generic class of AGN there are several types of sources: starburst galaxies  
(NGC 253 -- Abdo et al., 2010b), starburst/Seyfert 2 
(NGC 4945 -- Lenc  \& Tingay, 2009) 
a Low--Excitation FRI radio galaxy (PKS 0625--35 -- Gliozzi  et al., 2008) 
a High--Excitation FRI radio galaxy (Cen A), a Narrow--Line Seyfert 1
(PKS 2004--447 -- Abdo et al., 2009b). 
In the following plots  these sources
are labelled ``AGN" to distinguish them from the class of BL Lacs and FSRQ. We do not 
enter in the details of the properties of these sub--classes of AGNs 
because it is out of the scope of the present paper.

%----------------------------------------------------
\begin{figure}
%\hskip -0.3 cm
\psfig{figure=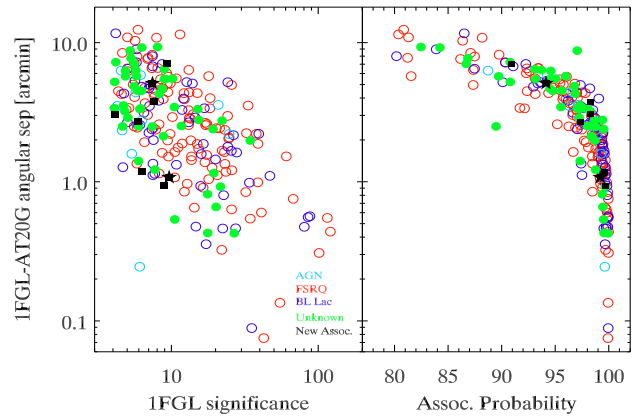,width=8.5cm,height=5.5cm}
\caption{
Angular separation between the centroid of 
the 1FGL source and the position of the associated AT20G source 
as a function of the \fe\ detection significance (left panel) 
and of the probability of the association (right panel) computed 
with the \texttt{gtsrcid} tool. 
Open circles are the associations (Tab. 1) which are also present in 
the 1LAC sample (i.e. already identified as blazars of different types: 
AGN, FRSQ, BL Lac). 
Those which are 1LAC likely blazars but of of unknown class are 
marked by filled (green) circles. 
The solid squares and stars
are the new associations (Tab. 2) found from 
the cross correlation of AT20G and 1FGL. 
The star symbols are the two new associations classified as FSRQ in this paper.}
\label{fg2}
\end{figure}
%----------------------------------------------------

Fig. \ref{fg2} shows the angular separation between 
the 1FGL and the AT20G sources for the 230 associations found 
as a function of the \fe\ detection significance (left panel) 
and of the posterior probability of the association (right panel). 
The angular separation between the 1FGL sources and the associated 
AT20G counterparts decreases as a function of the \fe\ detection significance. 
This behaviour is common among the different blazar types. 
The number of associations concentrates to low significance values 
where also the association separation is larger. From Fig. \ref{fg2} 
it appears also that, although our association methods selected only 
those with a posterior probability $>$80\%, most of the associations 
found have a probability distribution (right panel of Fig. \ref{fg2}) that is  
skewed towards high probabilities ($>95$\%) suggesting that most of 
\fe\ sources with relatively low detection significance are found 
in associations with a high posterior probability.

\section{Radio and gamma--ray properties of the AT20G--1FGL associations}

\begin{figure}
\hskip -0.5 cm
\psfig{figure=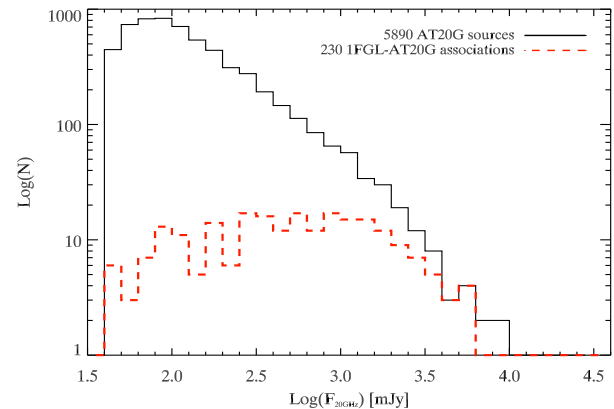,width=9.cm,height=6cm}
\caption{
Distribution
of the radio fluxes at 20 GHz for the 5890 sources of the AT20G 
sample (solid line) and of 230 sources associated with one \fe\ 
1FGL source with a probability $>80$\% of the association.
}
\label{fg3}
\end{figure}

In Fig. \ref{fg3} we compare the distribution of the 20 GHz fluxes of all 
the 5890 sources of the AT20G sample with the distribution of the radio 
flux of those sources associated with one source of the 1FGL catalogue. 
The 1FGL sources select the radio counterparts with the largest 
fluxes within the flux--limited  AT20G survey. 
The Kolmogorov--Smirnov (KS) test gives a probability of 1.9\% that the 
two distributions are similar. 

Selecting the associations that are blazars of known type we compare 
the distributions of radio fluxes of BL Lacs and FSRQs. 
In the LBAS sample, these were found (Abdo et al. 2009) to be different
(considering the 8.4 GHz fluxes).
Using our sample and the 20 GHz flux we also find that these two 
populations have different radio fluxes with FSRQ being on average 
brighter than BL Lacs (the chance probability that the two distribution 
are consistent being only 1.5$\times 10^{-4}$).

%----------------------------------------------------
%----------------------------------------------------

%----------------------------------------------------
\begin{figure*}
\hskip -0.7 cm
\psfig{figure=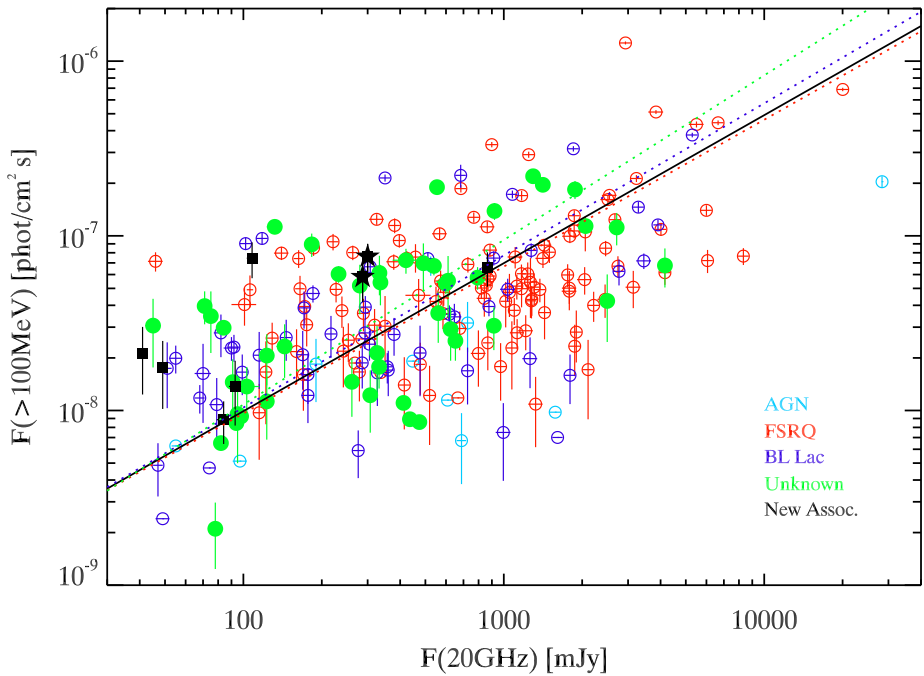,width=14cm,height=10cm}
\caption{
Correlation between the $\gamma$--ray flux (integrated above 100 MeV) 
and the radio flux at 20 GHz for the 230 associations found from the 
cross correlation of the AT20G survey and the 1FGL catalogue. 
Symbols are the same as in Fig. \ref{fg2}. 
The solid line is the best fit correlation of all the data points, 
the dotted line for FSRQ, the dashed line for BL Lac and the 
dot--dashed line for sources of unknown optical classification. 
The slopes and significance of the different correlations 
are reported in Tab. \ref{tab1}.  
}
\label{fg4}
\end{figure*}
%----------------------------------------------------

The radio vs $\gamma$--ray properties of the \fe\ first 
3 months catalogue (Abdo et al. 2009; Giroletti et al. 2010) 
showed that there is a possible correlation 
between the 8.4 GHz radio flux and the $\gamma$--ray flux. 
However, this correlation was found to be statistically significant 
(chance probability of 0.05\%) only for the population of BL Lacs, 
while FSRQ showed only a marginal correlation (with a chance 
probability of 8\%).
Without separating the source types, Kovalev (2009; 2009a)
% This correlation has been further investigated by Kovalev (2009; 2009a)
% with the first 3 months \fe\ survey through 111 sources without separating source types: they 
finds that the parsec--scale radio emission and the 
$\gamma$--ray flux are strongly related in the bright $\gamma$--ray 
objects of the LBAS sample. 
Similar results are found with the preliminary one--year \fe\ sample 
by Giroletti et al. (2010). 

Through our large sample of associations (230 sources) we can extend this 
correlation analysis in particular relying on the  lower limiting flux 
(40 mJy) at the higher frequency (i.e. 20 GHz) of the AT20G survey. 
Fig. \ref{fg4} shows that there is indeed a strong correlation between 
the radio flux at 20 GHz and the $\gamma$--ray flux (above 100 MeV) for 
the associations found from the cross correlation of the AT20G survey 
with the 1FGL/1LAC catalogue (\S 1). 

% ---------------------------------------------------
\begin{table}
\centering
\begin{tabular}{lccc}
\hline
\hline
Sources        &$r_{s}$ &$P$             &$A$  \\
\hline
All (230)        &0.53 &10$^{-17}$         & 0.85$\pm$0.04 \\
FSRQ (112)       &0.42 &3$\times10^{-6}$   & 0.84$\pm$0.06 \\
BL Lac (54)      &0.51 &9.4$\times10^{-6}$ & 0.87$\pm$0.07   \\
Unknown (46)     &0.58 &1.9$\times10^{-5}$ & 0.94$\pm$0.1   \\
\hline
\hline
\end{tabular}
\vskip 0.4 true cm
\caption{
Correlation between the $\gamma$--ray flux and the 20 GHz radio 
flux density for all 230 associations and for different sub--samples. 
The Spearman correlation coefficient $r_{s}$ and the chance correlation  probability 
$P$ are given. The correlation is fitted with a function 
$\log F_{>100\, {\rm MeV}} \propto A\,\log F_{20\, {\rm GHz}}$ and the slope 
is given in column 4.
}
\label{tab1}
\end{table}
% ---------------------------------------------------

Considering all the associations, the correlation has a rank correlation 
coefficient 0.53 and a chance probability of 10$^{-17}$. 
The best fit of the correlation (shown by the solid line in Fig. \ref{fg4}) obtained with the least 
square bisector method (e.g. Isobe et al. 1990)  
has a slope 0.85$\pm$0.04. 
Fig. \ref{fg4} shows that the population of blazars of unknown type 
(green filled circles) tends to have low radio fluxes, more similar to 
the distribution of BL Lacs (open blue circles in Fig. \ref{fg4}) in this plane. 
All AGNs (containing several source types -- see \S 2), lie below the 
correlation (cyan open circles in Fig. \ref{fg4}), having lower $\gamma$--ray 
fluxes compared to FSRQs and BL Lacs. 
Due to the different redshift spanned by FSRQs and BL Lacs it is worth 
(Abdo et al. 2009; Mucke et al. 1997) to investigate this correlation 
separately for the two classes. 
Correlation analysis results are given in Tab. \ref{tab1} where the 
Spearman correlation coefficient $r_{s}$ and its chance probability $P$ 
together with the correlation slope $A$ are given. 

We find significant correlations for FSRQs ($P=3\times 10^{-6}$),
for BL Lacs ($P=9.4\times 10^{-6}$) and also for the 
blazars of unknown optical classification ($P=1.9\times 10^{-5}$). 
The slopes of all these three correlations are 
% The correlations for the three sub--groups (FSRQ, BL Lacs and unknown blazars)  
% have slopes which are all 
consistent at the 1$\sigma$ level (last column in Tab. \ref{tab1}).  
Comparing to the results obtained from the LBAS sample (Abdo et al. 2009) 
we confirm the existence of a strong correlation between the radio and 
the $\gamma$--ray flux for BL Lacs while we find, for the first time, 
a similar strong correlation also for the population of FSRQs. 
The correlation found between the $\gamma$--ray and radio flux could be subject to possible biases 
related to the flux limits of the radio and $\gamma$--ray samples considered. Detailed study of possible selection effects 
is beyond the scope of the present work and requires numerical simulations. This will be the subject of a forthcoming paper. 
% Moreover, we note that the radio flux densities are not measured 
% simultaneously to the $\gamma$ ray flux. In Abdo et al. 2009 this 
% correlation was investigated for the LBAS sample considering the 
% peak $\gamma$ ray flux instead of the average one as we do here. 
% We note, however, that considering the peak flux of the 1FGL sources 
% in our association list we still find a strong correlation between 
% these two quantities, although for several sources the peak flux 
% in the $\gamma$--ray band is missing in the 1FGL catalog. 
% Moreover, considering that FSRQ are less variable in the 
% $\gamma$--ray band than BL Lacs (Vero?), the finding of a 
% statistically robust correlation between the radio and 
% $\gamma$--ray flux for this class of objects using mean 
% fluxes is reassuring. Simulations of the possible biases on 
% the correlation found are beyond the scope of the present 
% paper and will be presented elsewhere.  

\begin{figure}
\hskip -0.7 cm
\psfig{figure=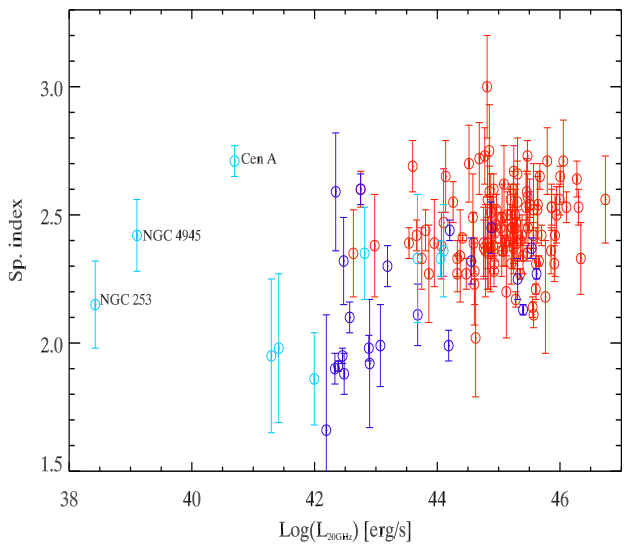,width=9.5cm,height=8cm}
\caption{
The photon $\gamma$--ray spectral index versus radio luminosity for 
the associations between the AT20G and the 1FGL sample which are 
also present in the 1LAC sample and have measured redshifts (144 sources). 
112 FSRQs, 22 BL Lacs, 10 AGNs  and 1 of unknown type are marked with red, blue and cyan symbols, 
respectively.}
\label{fg4a}
\end{figure}

For the 144 out of 230 associations found with sources with measured 
redshift we can also compare the distribution of FSRQ, BL Lac 
and AGN in the plane of the $\gamma$--ray photon spectral index versus the 
radio luminosity (at 20 GHz). 
All the 112 FSRQs have a known redshift , while only 
22 (out of 54) BL Lacs and all 10  sources classified AGNs  
have a known $z$.
This is shown in Fig. \ref{fg4a}.

It has been shown (Ghisellini, Tavecchio \& Maraschi 2009)
that FSRQs and BL Lacs tend to divide based on the $\gamma$--ray spectral 
index versus $\gamma$--ray luminosity. 
Here we see that also the in the spectral index -- radio luminosity 
plane BL Lacs and FSRQ occupy different regions, with FSRQs being
steeper and more luminous.
Among the sources generically classified as AGN in the 1LAC sample we have marked in 
Fig. \ref{fg4a} the starburst NGC 253, the starburst/Seyfer 2 NGC 4945 and the radio galaxy Cen A which have 
lower radio luminosity with respect to FSRQs and BL Lacs.
% having larger radio luminosity at 20 GHz than BL Lac. 
%Furthermore, as noted in Abdo et al. (2009), the non blazars AGNs 
%tend to be located in a different region with respect to blazars, 
%being softer and weaker.
%{\bf Ma nella figura ce ne sono solo 2 softer and weaker....
%Gli altri seguono i BL Lac... e mi sa che i due che sono fuori 
%sono due radiogalassie? Possiamo metterci i nomi? E se  devi rifare la figura,
%possiamo fare l'asse x in Log e la label Log $L_{20 \, \rm GHz}$ 
%cosi' i numeri vengono piu' grandi?!?..!?
%}

However, considering that the number of associations with measured redshifts 
is only a fraction of the associations found, this plot is still highly 
incomplete, particularly for the population of BL Lacs. 

%----------------------------------------------------
%----------------------------------------------------

\begin{figure}
\hskip -0.7 cm
\psfig{figure=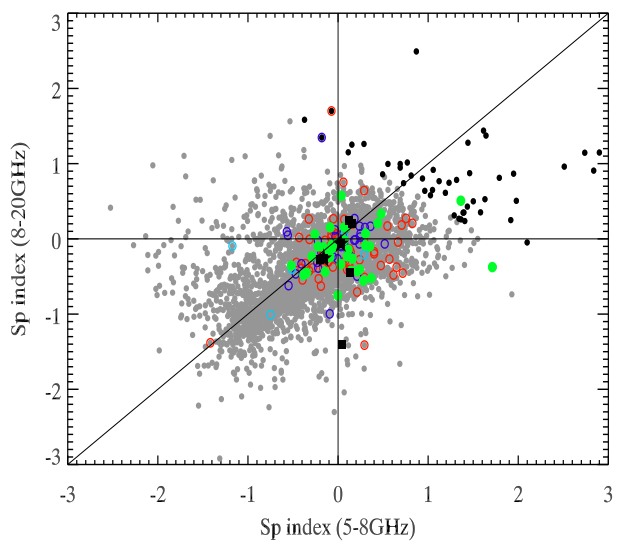,width=9.5cm,height=8cm}
\caption{
Low frequency (5--8 GHz) versus high frequency 
(8--20 GHz) spectral index for the 3795 AT20G sources with 
flux measurements at all the three frequencies. 
Black dots are the AT20G sources defined ``Ultra--Inverted Spectrum" (UIS) 
which have $\alpha(5-20{\, \rm GHz})>+0.7$ (see Murphy et al. 2009). 
The open and filled symbols (same as Fig. \ref{fg2}) represent 
the associations found from the cross correlation of the AT20G 
sample and the \fe\ 1FGL catalogue.}
\label{fg5}
\end{figure}

In the AT20G sample 3795 sources have radio fluxes measured at 20, 8 and 5 GHz. 
In most of these cases, moreover, the measurements are at the same epoch. 
From these fluxes we can estimate the radio spectral index as 
$\alpha_{1-2}=\log(S_{1}/S_{2})/\log(\nu_{1}/\nu_{2})$, 
where $S_{i=1,2}$ is the flux measured at a certain frequency $\nu_{i=1,2}$. 
The high frequency (8--20 GHz) versus the low frequency (5--8 GHz) 
spectral index for the AT20G sample (grey dots) and of the 1FGL--AT20G 
associations are shown in Fig. \ref{fg5}. 
All blazars in the 1LAC sample associated with a counterpart in the AT20G sample 
have flat radio spectra: the average 
$\langle\alpha(5-8{\, \rm GHz})\rangle=0.02\pm$0.28 (1$\sigma$) and 
$\langle\alpha(8-20{\, \rm GHz})\rangle=-0.16\pm$0.27 (1$\sigma$). 
There is no difference in the radio spectrum at low or high 
frequencies among the different blazar classes (FSRQ or BL Lac) 
and also the blazar candidates of unknown type (green filled circles 
in Fig. \ref{fg5}) have the same distribution in the 
$\alpha(5-8{\rm GHz})-\alpha(8-20{\rm GHz})$ plane. 
This is indicative that in these sources we are observing the 
partially self--absorbed emission from a compact radio core. 

The AT20G survey contains $\sim$1.2\% of sources with with extremely 
hard radio spectra, i.e. $\alpha(5-20{\rm GHz})>+0.7$ called 
``Ultra--Inverted Spectrum" (UIS) sources (black circles in Fig. \ref{fg5}). 
Among these sources we find 3 associations: one is the FSRQ 
PKS 0601--70 associated to 1FGL J0600.7--7037 and one is the BL Lac 
CRATES J1918--4111 associated to 1FGL J1918.4--4108. 
The third source, of the unknown class (i.e. the green circle among the 
UIS sources in Fig. \ref{fg5}), is PMN 1326--5256,
associated with 1FGL J1327.0--5257. 
This source is particularly interesting, being the brightest 
radio source in the sub-sample of UIS. 
This source and its SED will be presented in Sec. 4. 
Fig. \ref{fg5} shows that the sources detected by \fe\ with a 
radio counterparts are not preferentially of the UIS type.
In other words, being UIS does not give an extra chance to be a strong
$\gamma$--ray blazar.

%----------------------------------------------------
%----------------------------------------------------

\section{Spectral classification of new blazar candidates}

From the cross correlation of the AT20G and the 1FGL catalogues we found 
46 associations which are classified as blazars (i.e. present in the 
1LAC sample) but of unknown type. 
For these sources it is important to find some possible evidences that might lead 
to their classification as either BL Lac or FSRQ.  
Similarly we have found 8 new associations. 
Several diagnostics are used for the classification of blazars, 
the main one relies on their optical spectra. 
Different classification criteria exist: in the 1LAC sample a BL Lac should 
show no or weak lines  
(equivalent width $<$ 5 \AA) and the Ca II H/K ratio $<$0.4 (A10a). 

Another approach is to build the broad band SED and compare it with those of 
the two classes (FSRQ and BL Lac) to possibly identify similar signatures. 
We also model the SED, adopting the modelling of Ghisellini et al. (2009).

An on--going project is to study the 46 associations of unknown 
type in order to classify them. 
Here we present, as a first result, one of these sources, 1FGL 1327.0--5257 
(associated to PMN J1326--5256 and to the AT20G J132649--525623) which is 
the source with the highest radio flux density of the UIS class. 

Among the 8 new associations we searched for any \sw\ pointing containing 
these sources. 
In two cases (1FGL J0904.7--3514 and 1FGL J1656.2--3257) there are 
\sw\ observations of the field. We associate 1FGL J1656.2--3257 
with the Swift J1656.3--3302 already classified as a FSRQ at $z\sim$2.4 
(Masetti et al. 2008) and we present here its broad band SED 
by combining the published data, the XRT data and the Fermi ones. 
We also build the SED of 1FGL J0904.7--3514 and we classify it as a FSRQ.

The \sw\ data were analysed with the software distributed as part of the 
\texttt{Heasoft v. 6.7} and the calibration database updated to 
September 2009 was adopted. 
The XRT data were processed with the standard procedure. 
We considered photon counting (PC) mode data with the standard 0--12 grade selection. 
Source events were extracted in a circular region of  aperture $\sim 47''$, 
and background was estimated in a same sized circular region far from the source. 
Response matrices were created through the \texttt{xrtmkarf} task. 
The channels with energies below 0.3 keV and above 10 keV were excluded from the fit. 
The spectra were analysed through XSPEC({\it v11.3.2ag}) with an absorbed power law 
with a fixed Galactic column density (Kalberla et al. 2005). 

UVOT (Roming et al. 2005) source counts were extracted from a 
circular region $5''$--sized centred on the source position, 
while the background was extracted from a larger nearby source-free region. 
Data were integrated with the \texttt{uvotimsum} task and then analysed 
through the \texttt{uvotsource} task. The observed magnitudes have been 
dereddened according to the formulae by Cardelli et al. (1989) and converted 
into fluxes by using standard formulae and zero points from Poole et al. (2008).

\subsubsection{1FGL J0904.7--3514}

The \fe\ source centroid is at RA=136.195$^\circ$ and Dec=--35.248$^\circ$ 
with a 68\% uncertainty radius (almost circular) of 0.06$^\circ$. 
There are three \sw\ pointings containing this position: 
one performed on 15--06--2009 ($\sim$5 ks) and two in January 2010 
(of $\sim$6 and 7 ks). 
The \fe\ 1FGL J0904.7--3514 source has a $\gamma$--ray flux (averaged over 
the 11 months survey and integrated above 100 MeV) of 7.6$\times 10^{-8}$ 
phot cm$^{-2}$ s$^{-1}$ and its photon spectral index is 2.69$\pm$0.1. 
Our cross correlation of the AT20G catalogue with the 1FGL associated this source 
with AT20G J090442--351423, which is $\sim$1 arcmin distant from the \fe\ centroid. 

\begin{figure}
\hskip 0.5 cm
\psfig{figure=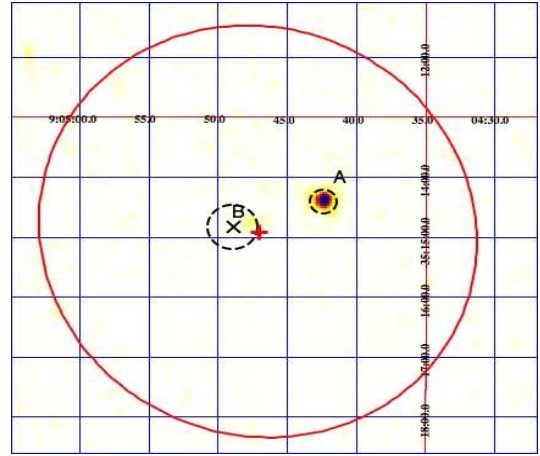,width=7cm,height=6cm}
\caption{
XRT image of 1FGL J0904.7--3514. 
The \fe/LAT centroid (marked with the plus sign) and its 68\% error 
ellipse is shown (solid red line). 
The bright X-ray source (labelled A)is coincident with the 
position of the radio AT20G source AT20GJ090442--35142 
(marked by the black dashed circle). 
The position of the other source within the field, PMN J0904--3514 
(labelled B) is shown by the dashed ellipse.}
\label{fg6}
\end{figure}

By summing all the \sw\ XRT images (Fig. \ref{fg6}) we find within 
the 68\% error ellipse of the \fe\ source a bright X--ray source 
coincident with AT20GJ090442--35142 (source A  in Fig. \ref{fg6}). 
By searching in NED for the possible other sources present in this 
field we found that there is PMN J0904--3514 that was one of the possible 
5 counterparts of the source 3EGJ0903--3531 (Mattox et al. 2001) and it 
is closer to the \fe\ centroid (source B - dashed ellipse in Fig. \ref{fg6}). 
However, the possible X--ray counterpart of source B (which is 
located at the border of its error ellipse) is a factor 10 fainter 
than source A which we claim is the X--ray and coincident 
radio counterpart of the \fe\ 1FGL J0904.7--3514.

%----------------------------------------------------
%----------------------------------------------------

\begin{figure}
\hskip -0.3 cm
\psfig{figure=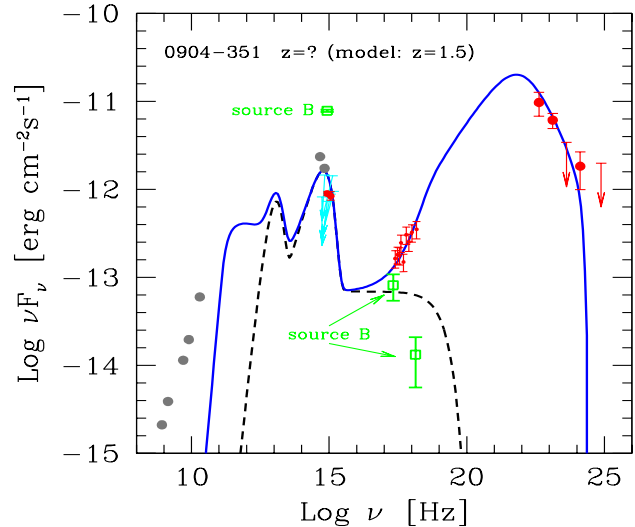,width=9cm,height=8cm}
\caption{
SED of 1FGL J0904.7--3514. Red points are XRT and UVOT 
data and at high energies the \fe\ LAT data taken from the 1FGL catalogue. 
The green points are the XRT and UVOT data of the other source within 
the \fe\ 68\% confidence ellipse (which is not present in the AT20G sample). 
Radio data are from the AT20G survey (at 5, 8 and 20 GHz) and other 
data are collected from the literature. 
}
\label{fg7}
\end{figure}

In Fig. \ref{fg7} we show the SED of 1FGL J0904.7--3514 obtained combining 
the \fe\ data and the X--ray and Optical/UV data available (red symbols). 
For comparison we also show (green symbols) the X--ray and optical data 
of source B in the field, i.e. PMN J0904--3514. 
We see that the SED of the association 1FGL J0904.7--3514 
with AT20G J090442--35142 (source A)
can be well modelled as a typical blazar with a model redshift z$\sim$1.5
while the other source has a very steep X--ray spectrum inconsistent with 
the \fe\ data.

%----------------------------------------------------
%----------------------------------------------------

\subsubsection{1FGL J1656.2--3257}

\begin{figure}
\hskip -0.3 cm
\psfig{figure=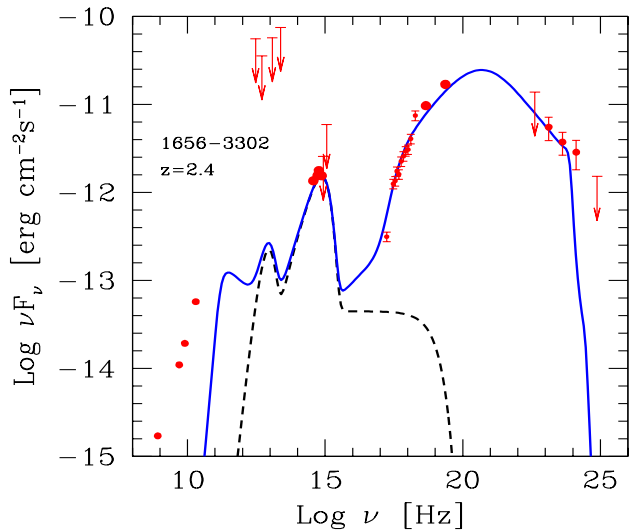,width=9cm,height=8cm}
\caption{
SED of 1FGL J1656.2--3257. 
}
\label{fg8}
\end{figure}

The \fe\ source centroid is at RA=254.055 and Dec=--32.9513 with a 
68\% confidence ellipse almost circular with radius 0.07$^\circ$. 
Within this region there is AT20G J165616--330207 which is at a 
separation of 0.085$^\circ$ and has an association probability of 94\%. 
The \fe\ 1FGL J1656.2--3257 has a photon spectral index 
$\Gamma$=2.43$\pm$0.11 and a $\gamma$--ray flux 
(averaged over the 11 months survey and integrated above 100 MeV) 
of 5.81$\times 10^{-8}$ phot cm$^{-2}$ s$^{-1}$. 
The position of the radio counterpart AT20G J165616--330207 that 
we associate with the \fe\ one is coincident with the \sw\ 
source SWIFT J1656--3302 (Bird et al. 2007; Malizia et al. 2007; 
Krivonos et al. 2007; Bodaghee et al. 2007; Sazonov et al. 2007). This source 
was classified as a FSRQ at redshift $z\sim2.4$ (Masetti et al. 2008).
There are two \sw\ pointings of SWIFT J1656--3302 in June 2006 (with an exposure of  
$\sim$4.3 ks and 4.8 ks).
We used these observations to build the SED  
shown in Fig. \ref{fg8}, collecting the other data from Masetti et al. (2008). 
Also in this case the source appears to be well 
modelled by the typical SED of a FSRQ. 

%----------------------------------------------------
%----------------------------------------------------

\subsubsection{1FGL J1327--5257}

The \fe\ centroid is at RA=201.75$^\circ$ and Dec=--52.96$^\circ$ 
with an error radius of 0.04$^\circ$. 
We associate this source to the AT20G J132649--525623. 
The \fe\ source is already associated in the 1LAC sample to PMN J1326--5256 
which is coincident with the AT20G source that we associated with the 1FGL one. 
The $\gamma$--ray flux of the \fe\ source, integrated above 100 MeV 
is 6.6$\times10^{-8}$ phot cm$^{-2}$ s$^{-1}$ and the photon spectral index is 2.32$\pm$0.06. 

PMN J1326--5256 is a compact radio source with scarce data in the literature.  
The AT20G counterpart is the brightest source at 20 GHz among the class 
of ``Ultra--Inverted Spectrum" (UIS) in this survey. 
It is detected at a  flux level of 2.061$\pm$0.1, 1.350$\pm$0.07 
and 0.606$\pm$0.03 Jy at 20 GHz, 8 GHz and 5 GHz, respectively, 
with a spectral index $\alpha$(5, 20 GHz)=+0.8$\pm$0.07 ($F(\nu)\propto \nu^\alpha$).

This source is intriguing in many aspects.  
Its redshift is unknown but its classification as a FSRQ seems to be excluded by Bignall et al. (2008) 
based on its featureless spectrum (in the 5000 -- 9100 \AA\ band).

The source has no large scale structure at 2.3 GHz, being unresolved at 
the level of 16 mas resolution of ATCA and AT--LBA (Bingall et al. 2008). 
Its radio properties have been extensively monitored 
(Cim\'o et al. 2006) because of its intra--day variability. 

\begin{center}
\begin{figure}
%\vskip -0.2 cm
\psfig{figure=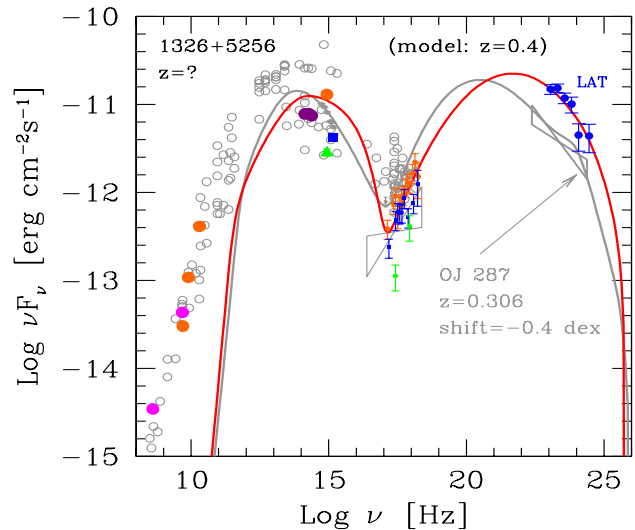,width=9.cm,height=8cm}
%\vskip -0.6  cm
\caption{
SED of J1326--5256 (coloured symbols). The radio data at 408 MHz and 4.85 GHz 
are shown with the magenta filled circles, while the AT20G fluxes 
(at 5, 8 and 20 GHz) are show by the orange circles. 
2MASS J, H, K data are shown with the black symbols. 
The red, blue, green symbols correspond to the three \sw\ observations. 
The open grey circles and grey bows represent the data of OJ287 
(from Ghisellini et al. 2009, Tavecchio et al. 2010) and the grey 
solid line is the best fit model, scaled in flux 
by a factor 0.4 dex. 
The solid blue line is the best fit model to SED of J1326--5256.  }
\label{fg9}
\end{figure}
\end{center}

There are three \sw\ observations of  J1326--5256: two in August 2009 
and one in October 2009.  
Fig. \ref{fg9} shows the SED of 
J1326--5256 with all the available data (coloured symbols). 
For comparison we also report the data of the classic 
BL Lac OJ 287 (grey symbols) shifted in flux by a factor 0.4 dex. 
The two models (blue and grey solid lines) represent the fit with 
the model described in Ghisellini \& Tavecchio (2009). %(MNRAS, 399, 2041). 
The close similarity of the two SEDs leads us to propose that 
J1326--5256 is a BL Lac. 
If it is slightly weaker than OJ 287 (that has $z=0.306$) 
because it is more distant, 
than the redshift of J1326--5256 is $z\sim 0.4$.

%--------------------------------------------------
%--------------------------------------------------

\section{Conclusions}

In this paper we cross correlated the \fe\ 1FGL catalog (A10) 
containing the sources detected above 
100 MeV during the 11 months survey with a complete sample of radio sources selected by the 
AT20G survey in the southern emisphere with 20 GHz flux density 
larger than 40 mJy. 
The cross correlation led to identify highly probable (association 
probability $>$80\%) radio counterparts for 230 1FGL sources. 
222 of these are already classified as blazars in the 1LAC catalog 
(A10a) and 8 sources are new associations.

Using the sample of the associated sources
and considering the uniform radio flux measurement of the AT20G survey, we have studied the 
radio to $\gamma$--ray flux correlation among 
different blazar sub--classes
finding that there is a significant correlation between these 
two fluxes both for BL Lacs and for FSRQs. 
Such a correlation has a slope of $\sim$0.85 for both blazars sub--classes.
% and it is consistent at 1 $\sigma$ between the two classes of blazars. 
If this correlation is further confirmed in the future by the extension of the sample 
of \fe\ blazars with measured radio fluxes, it would help to 
correctly estimate the contribution of blazars
to the $\gamma$--ray background (e.g. Stecker et al. 1993).
% shed light on the fraction of the unresolved $\gamma$--ray background (e.g. REF). 

The radio properties of the 230 associations show also that they are typically 
flat spectrum radio sources. 
In particular, steep spectrum radio sources are not the radio 
counterparts to \fe\ sources, but also radio sources with extremely hard radio spectra (i.e. 
$\alpha$(5--20 GHz)$>$+0.7) which are the 1.2\% of the AT20G sample are 
not preferentially detected by \fe. 
Indeed, we find only three associations with such hard radio sources. 

We have presented the spectral energy distribution of three sources: 
two (1FGL J0904.7--3514 and 1FGL J1656.2--3257) are among the eight 
new associations that we have found by cross correlating the \fe\ 
sample and the AT20G survey. 
Their spectrum, obtained by combining radio, \sw\ and \fe\ data resembles that of a
typical FSRQ. 
The third source (1FGL J1327--5257) is presented as a first 
interesting result of an on--going program aimed at studying 
and classify the  unknown blazar candidates of the 1LAC sample. 
This source % (1FGL J1327--5257) 
is among the brightest radio sources in the AT20G survey and it is 
characterised by a rapidly variable radio emission. 
Its SED is remarkably similar to that of the prototypical BL Lac OJ 287, but located
at the slightly greater redshift $z\sim$0.4. 

% --------------------------------------------------------
\begin{table} 
\centering
\begin{tabular}{|l|l|l|l|l|l|}
\hline
\hline
\small
%  \multicolumn{1}{|l|}{RA} &
%  \multicolumn{1}{l|}{Dec} &
%  \multicolumn{1}{l|}{P} &
%  \multicolumn{1}{l|}{Sep} &
%  \multicolumn{1}{l|}{1FGL} &
%  \multicolumn{1}{l|}{AT20G} \\
  RA      &Dec      &P    &Sep    &1FGL              &AT20G          \\
  deg     &deg      &     &arcmin &                  &               \\
\hline
  116.613 &--07.164 &0.97 &2.70 &J0746.5--0711       &J074627--070951 \\
  136.176 &--35.239 &0.99 &1.07 &{\bf J0904.7--3514} &J090442--351423 \\
  188.529 &--57.598 &0.99 &1.18 &J1234.0--5736       &J123407--573552 \\
  194.020 &--59.328 &0.98 &3.06 &J1256.1--5922       &J125604--591943 \\
  198.767 &--53.576 &0.98 &3.77 &J1314.9--5338       &J131504--533436 \\
  254.070 &--33.035 &0.94 &5.10 &{\bf J1656.2--3257} &J165616--330207 \\
  262.945 &--30.052 &0.90 &7.07 &J1732.0--2957       &J173146--300309 \\
  275.910 &--34.903 &0.99 &0.93 &J1823.5--3454       &J182338--345412 \\
\hline
\hline
\end{tabular}
% \vskip 0.4 true cm
\caption{
The eight new associations found by the cross-correlation of the 
1FGL sample and the AT20G survey. 
Positions (RA and Dec) of the associated radio source, probability of 
the association and angular separation between the AT20G counterpart 
and the 1FGL centroid are reported. 
The two sources (J0904 classified as FSRQ in this paper) whose SED is studied in this work are marked in boldface. }
\label{tab2}
\end{table}

\section*{Acknowledgments}
This work was partly financially supported by a 2007 COFIN-MIUR grant and 
ASI I/088/06/0 grant. This research has made use of the NASA/IPAC Extragalactic 
Database (NED) which is operated by the Jet Propulsion Laboratory, 
California Institute of Technology, under contract with the 
National Aeronautics and Space Administration.
This work is based on the public available data of the \sw\ satellite
and \fe\ LAT instrument obtained through the HEASARC.

\onecolumn
\newpage
%-----------------------------------------------------------------------------------------------------------------------------------
\begin{center}
\small
\begin{longtable}{rrrrlllcc}
%\begin{table*} 
%\centering
%\begin{tabular}{|r|r|r|r|l|l|l|l|l|}
\hline \hline
  \multicolumn{1}{r|}{RA} &
  \multicolumn{1}{r|}{Dec} &
  \multicolumn{1}{c|}{P} &
  \multicolumn{1}{c|}{Sep} &
  \multicolumn{1}{l|}{AT20G} &
  \multicolumn{1}{l|}{1FGL} &
  \multicolumn{1}{l|}{1LAC} &
  \multicolumn{1}{c|}{Class} &
  \multicolumn{1}{c}{Clean} \\
  deg    &   deg    &          & arcmin &                           &                          &                                    &        &  \\
\hline
\endhead
  0.325 & -7.774 & 0.94   & 5.32 &  J000118-074626 &   J0000.9-0745 & CRATES J0001-0746 & BLL & Y\\
  1.148 & -47.605 & 0.97  & 1.92 &  J000435-473619 &   J0004.7-4737 & PKS 0002-478 & FSRQ & Y\\
  3.249 & -39.907 & 0.98  & 2.67 &  J001259-395426 &   J0013.1-3952 & PKS 0010-401 & BLL & Y\\
  4.399 & -5.2115 & 0.98  & 2.42 &  J001735-051241 &   J0017.4-0510 & CGRABS J0017-0512 & FSRQ & Y\\
  7.572 & -42.412 & 0.94  & 4.72 &  J003017-422446 &   J0029.9-4221 & PKS 0027-426 & FSRQ & Y\\
  9.561 & -24.983 & 0.92  & 6.03 &  J003814-245901 &   J0038.4-2504 & PKS 0035-252 & FSRQ & Y\\
  11.887 & -25.288 & 0.94 & 5.84 &  J004733-251717 &   J0047.3-2512 & NGC 253 & AGN$^{(a)}$ & Y\\
  12.497 & -57.641 & 0.99 & 1.14 &  J004959-573827 &   J0049.8-5738 & PKS 0047-579 & FSRQ & Y\\
  12.589 & -4.872 & 0.90  & 7.17 &  J005021-045221 &   J0050.0-0446 & PKS 0047-051 & FSRQ & Y\\
  12.672 & -9.484 & 0.99  & 0.46 &  J005041-092905 &   J0050.6-0928 & PKS 0048-09 & BLL & Y\\
  12.784 & -6.834 & 0.99  & 0.96 &  J005108-065004 &   J0051.1-0649 & PKS 0048-071 & FSRQ & Y\\
  14.509 & -32.572 & 0.98 & 4.69 &  J005802-323420 &   J0058.4-3235 & PKS 0055-328 & BLL & Y\\
  19.052 & -11.604 & 0.81 & 10.9 &  J011612-113614 &   J0115.5-1132 & PKS 0113-118 & FSRQ & Y\\
  19.738 & -21.691 & 0.87 & 4.96 &  J011857-214130 &   J0118.7-2137 & PKS 0116-219 & FSRQ & Y\\
  20.132 & -27.023 & 0.99 & 0.46 &  J012031-270124 &   J0120.5-2700 & PKS 0118-272 & BLL & Y\\
  23.181 & -16.913 & 0.99 & 1.34 &  J013243-165448 &   J0132.6-1655 & PKS 0130-17 & FSRQ & Y\\
  24.409 & -24.514 & 0.98 & 2.20 &  J013738-243053 &   J0137.5-2428 & PKS 0135-247 & FSRQ & Y\\
  25.407 & -9.500 & 0.99  & 1.32 &  J014137-093001 &   J0141.7-0929 & PKS 0139-09 & BLL & Y\\
  26.264 & -27.559 & 0.98 & 2.04 &  J014503-273333 &   J0144.9-2732 & PKS 0142-278 & FSRQ & Y\\
  29.463 & -46.239 & 0.96 & 3.67 &  J015751-461423 &   J0157.5-4613 & CGRABS J0157-4614 & FSRQ & Y\\
  29.658 & -39.534 & 0.92 & 7.02 &  J015838-393204 &   J0158.0-3931 & CGRABS J0158-3932 & BLL & Y\\
  29.930 & -27.677 & 0.97 & 1.11 &  J015943-274038 &   J0159.7-2741 & CRATES J0159-2740 & BLL & Y\\
  31.240 & -17.022 & 0.98 & 1.52 &  J020457-170120 &   J0205.0-1702 & PKS 0202-17 & FSRQ & Y\\
  32.692 & -51.017 & 0.99 & 0.98 &  J021046-510101 &   J0210.6-5101 & PKS 0208-512 & BLL & Y\\
  34.261 & -8.347 & 0.87  & 8.43 &  J021702-082052 &   J0217.0-0829 & PKS 0214-085 & FSRQ & Y\\
  35.503 & -16.254 & 0.95 & 4.09 &  J022200-161516 &   J0222.1-1618 & PKS 0219-164 & FSRQ & Y\\
  37.368 & -36.732 & 0.99 & 1.26 &  J022928-364356 &   J0229.3-3644 & PKS 0227-369 & FSRQ & Y\\
  40.669 & -0.012 & 0.97  & 8.75 &  J024240-000046 &   J0242.7+0007 & RX J0241.9+0009 & UNKNOWN & N\\
  41.500 & -46.854 & 0.99 & 1.17 &  J024600-465116 &   J0245.9-4652 & PKS 0244-470 & FSRQ & Y\\
  43.199 & -22.323 & 0.99 & 0.54 &  J025247-221924 &   J0252.8-2219 & PKS 0250-225 & FSRQ & Y\\
  44.420 & -12.200 & 0.89 & 8.19 &  J025740-121201 &   J0257.8-1204 & CGRABS J0257-1212 & FSRQ & Y\\
  45.961 & -62.190 & 0.97 & 2.91 &  J030350-621125 &   J0303.4-6209 & PKS 0302-623 & FSRQ & Y\\
  45.860 & -24.119 & 0.99 & 1.62 &  J030326-240711 &   J0303.5-2406 & PKS 0301-243 & BLL & Y\\
  47.483 & -60.977 & 0.98 & 1.51 &  J030956-605839 &   J0310.1-6058 & PKS 0308-611 & FSRQ & Y\\
  48.987 & -10.527 & 0.99 & 1.61 &  J031556-103138 &   J0315.9-1033 & PKS 0313-107 & FSRQ & Y\\
  51.344 & -56.484 & 0.97 & 3.33 &  J032522-562905 &   J0325.6-5626 & CRATES J0325-5635 & UNKNOWN & N\\
  53.556 & -40.140 & 0.99 & 1.66 &  J033413-400825 &   J0334.2-4010 & PKS 0332-403 & BLL & Y\\
  53.564 & -37.428 & 0.98 & 2.89 &  J033415-372543 &   J0334.4-3727 & CRATES J0334-3725 & BLL & Y\\
  54.807 & -17.600 & 0.99 & 1.58 &  J033913-173600 &   J0339.1-1734 & PKS 0336-177 & AGN & Y\\
  54.878 & -1.776 & 0.96  & 5.69 &  J033930-014635 &   J0339.2-0143 & PKS 0336-01 & FSRQ & N\\
  55.830 & -25.504 & 0.92 & 6.23 &  J034319-253017 &   J0343.4-2536 & PKS 0341-256 & FSRQ & Y\\
  57.158 & -27.820 & 0.96 & 2.44 &  J034838-274913 &   J0348.5-2751 & PKS 0346-279 & FSRQ & Y\\
  57.490 & -21.046 & 0.99 & 2.08 &  J034957-210247 &   J0349.9-2104 & PKS 0347-211 & FSRQ & Y\\
  59.251 & -49.929 & 0.87 & 6.49 &  J035700-495547 &   J0357.1-4949 & PKS 0355-500 & BLL & Y\\
  60.503 & -26.261 & 0.97 & 3.37 &  J040200-261540 &   J0402.1-2618 & CRATES J0402-2615 & UNKNOWN & Y\\
  60.974 & -36.083 & 0.99 & 1.21 &  J040353-360500 &   J0403.9-3603 & PKS 0402-362 & FSRQ & Y\\
  61.391 & -13.137 & 0.98 & 2.33 &  J040534-130813 &   J0405.6-1309 & 1WGA J0405.6-1313 & AGN & N\\
  61.745 & -38.440 & 0.81 & 5.74 &  J040658-382627 &   J0407.4-3827 & PKS 0405-385 & FSRQ & Y\\
  63.305 & -53.533 & 0.92 & 3.37 &  J041313-533200 &   J0413.4-5334 & CRATES J0413-5332 & FSRQ & Y\\
  64.152 & -18.852 & 0.98 & 0.84 &  J041636-185108 &   J0416.5-1851 & PKS 0414-189 & FSRQ & Y\\
  65.544 & -6.729 & 0.96  & 4.06 &  J042210-064344 &   J0422.0-0647 & CRATES J0422-0643 & FSRQ & Y\\
  65.815 & -1.342 & 0.99  & 2.23 &  J042315-012033 &   J0423.2-0118 & PKS 0420-01 & FSRQ & Y\\
  67.168 & -37.938 & 0.99 & 0.56 &  J042840-375619 &   J0428.6-3756 & PKS 0426-380 & BLL & Y\\
  68.533 & -20.254 & 0.96 & 3.13 &  J043408-201517 &   J0434.1-2018 & CRATES J0434-2015 & BLL & Y\\
  69.645 & -12.851 & 0.97 & 3.29 &  J043834-125103 &   J0438.8-1250 & PKS 0436-129 & FSRQ & Y\\
  70.660 & -0.295 & 0.99  & 2.20 &  J044238-001744 &   J0442.7-0019 & BZB J0442-0018 & BLL & N\\
  71.256 & -60.250 & 0.88 & 6.29 &  J044501-601500 &   J0445.2-6008 &	J0445-6015 & AGN & N\\
  72.353 & -43.835 & 0.99 & 1.10 &  J044924-435008 &   J0449.5-4350 & PKS 0447-439 & BLL & Y\\
  73.311 & -28.127 & 0.98 & 2.42 &  J045314-280737 &   J0453.2-2805 & PKS 0451-28 & FSRQ & Y\\
  73.961 & -46.266 & 0.98 & 3.06 &  J045550-461558 &   J0455.6-4618 & PKS 0454-46 & FSRQ & Y\\
  74.152 & -31.603 & 0.97 & 4.33 &  J045636-313611 &   J0456.4-3132 & CRATES J0456-3136 & FSRQ & Y\\
  74.263 & -23.414 & 0.99 & 0.54 &  J045703-232451 &   J0457.0-2325 & PKS 0454-234 & FSRQ & Y\\
  75.303 & -1.987 & 0.97  & 1.98 &  J050112-015914 &   J0501.0-0200 & PKS 0458-02 & FSRQ & Y\\
  76.463 & -4.323 & 0.91  & 3.39 &  J050551-041926 &   J0505.8-0416 & CRATES J0505-0419 & FSRQ & Y\\
  76.977 & -61.078 & 0.94 & 4.33 &  J050754-610442 &   J0507.3-6103 & PKS 0506-61 & FSRQ & N\\
  79.187 & -62.118 & 0.99 & 0.81 &  J051644-620706 &   J0516.7-6207 & PKS 0516-621 & UNKNOWN & Y\\
  80.741 & -36.458 & 0.90 & 5.19 &  J052257-362730 &   J0522.8-3632 & PKS 0521-36 & BLL & Y\\
  81.569 & -48.510 & 0.98 & 1.76 &  J052616-483036 &   J0526.3-4829 & PKS 0524-485 & FSRQ & Y\\
  83.418 & -83.410 & 0.99 & 1.07 &  J053340-832439 &   J0533.0-8324 & PKS 0541-834 & FSRQ & Y\\
  84.709 & -44.085 & 0.99 & 0.47 &  J053850-440508 &   J0538.8-4404 & PKS 0537-441 & BLL & Y\\
  84.975 & -28.665 & 0.80 & 12.3 &  J053954-283956 &   J0539.1-2847 & PKS 0537-286 & FSRQ & Y\\
  84.814 & -3.949 & 0.90  & 5.22 &  J053915-035657 &   J0539.4-0400 & CRATES J0539-0356 & UNKNOWN & N\\
  89.526 & -38.641 & 0.83 & 8.94 &  J055806-383830 &   J0557.6-3831 & CRATES J0558-3838 & BLL & Y\\
  89.693 & -74.985 & 0.97 & 2.38 &  J055846-745906 &   J0559.2-7500 & PKS 0600-749 & BLL & Y\\
  90.297 & -70.602 & 0.95 & 2.59 &  J060111-703609 &   J0600.7-7037 & PKS 0601-70 & FSRQ & Y\\
  92.006 & -15.343 & 0.99 & 0.65 &  J060801-152036 &   J0608.0-1521 & CRATES J0608-1520 & UNKNOWN & Y\\
  91.942 & -6.385 & 0.82  & 9.30 &  J060746-062307 &   J0608.1-0630c & CRATES J0609-0615 & UNKNOWN & N\\
  91.998 & -8.580 & 0.96  & 3.93 &  J060759-083449 &   J0608.2-0837 & PKS 0605-08 & FSRQ & Y\\
  94.389 & -17.256 & 0.96 & 3.82 &  J061733-171525 &   J0617.7-1718 & CRATES J0617-1715 & BLL & Y\\
  96.703 & -54.537 & 0.87 & 7.94 &  J062648-543214 &   J0625.9-5430 & CGRABS J0625-5438 & FSRQ & Y\\
  96.533 & -42.892 & 0.94 & 6.26 &  J062607-425332 &   J0626.6-4254 & CRATES J0626-4253 & UNKNOWN & Y\\
  96.778 & -35.487 & 0.99 & 2.76 &  J062706-352916 &   J0627.3-3530 & PKS 0625-35 & AGN$^{(c)}$ & Y\\
  97.349 & -19.988 & 0.97 & 3.50 &  J062923-195919 &   J0629.6-2000 & PKS 0627-199 & BLL & Y\\
  97.748 & -24.112 & 0.99 & 0.35 &  J063059-240645 &   J0630.9-2406 & CRATES J0630-2406 & BLL & Y\\
  98.943 & -75.271 & 0.95 & 5.09 &  J063546-751616 &   J0636.1-7521 & PKS 0637-75 & FSRQ & Y\\
  102.118 & -17.734 & 0.93& 5.53 &  J064828-174405 &   J0648.7-1740 & TXS 0646-176 & FSRQ & N\\
  102.602 & -16.627 & 0.95& 4.52 &  J065024-163739 &   J0650.6-1635 & PKS 0648-16 & UNKNOWN & N\\
  105.129 & -66.179 & 0.99& 0.92 &  J070031-661045 &   J0700.4-6611 & PKS 0700-661 & UNKNOWN & Y\\
  105.393 & -46.576 & 0.81& 7.77 &  J070134-463436 &   J0702.0-4628 & PKS 0700-465 & FSRQ & Y\\
  105.678 & -19.855 & 0.90& 7.24 &  J070242-195121 &   J0702.2-1954 & TXS 0700-197 & UNKNOWN & N\\
  111.460 & -0.915 & 0.98 & 2.56 &  J072550-005457 &   J0725.9-0053 & PKS 0723-008 & BLL & N\\
  112.58 & -11.687 & 0.99 & 0.75 &  J073019-114113 &   J0730.3-1141 & PKS 0727-11 & FSRQ & N\\
  113.680 & -77.187 & 0.94& 4.66 &  J073443-771114 &   J0734.1-7715 & PKS 0736-770 & UNKNOWN & Y\\
  118.610 & -11.787 & 0.99& 0.53 &  J075426-114716 &   J0754.4-1147 & OI -187 & UNKNOWN & N\\
  121.790 & -5.687 & 0.95 & 3.44 &  J080709-054115 &   J0807.0-0544 & PKS 0804-05 & UNKNOWN & Y\\
  122.064 & -7.852 & 0.99 & 0.99 &  J080815-075110 &   J0808.2-0750 & PKS 0805-07 & FSRQ & Y\\
  122.763 & -75.507 & 0.97& 2.51 &  J081103-753027 &   J0811.1-7527 & CRATES J0811-7530 & UNKNOWN & Y\\
  123.548 & -10.203 & 0.89& 5.72 &  J081411-101210 &   J0814.5-1011 &  AT20G J0814-1012 & UNKNOWN & N\\
  124.457 & -9.558 & 0.91 & 6.75 &  J081749-093330 &   J0818.0-0938 & CGRABS J0817-0933 & BLL & Y\\
  126.506 & -22.507 & 0.98& 2.16 &  J082601-223027 &   J0825.8-2230 & PKS 0823-223 & BLL & N\\
  126.464 & -32.306 & 0.98& 2.12 &  J082551-321823 &   J0825.9-3216 & PKS 0823-321 & UNKNOWN & N\\
  131.26 & -54.969 & 0.98 & 1.22 &  J084502-545808 &   J0845.0-5459 & PMN J0845-5458 & UNKNOWN & N\\
  131.756 & -23.617 & 0.98& 2.50 &  J084701-233701 &   J0846.9-2334 & CRATES J0847-2337 & UNKNOWN & Y\\
  132.440 & -35.683 & 0.97& 1.40 &  J084945-354101 &   J0849.6-3540 & VCS2 J0849-3541 & UNKNOWN & N\\
  132.539 & -12.226 & 0.99& 1.38 &  J085009-121337 &   J0850.0-1213 & CGRABS J0850-1213 & FSRQ & Y\\
  134.173 & -11.087 & 0.99& 0.42 &  J085641-110514 &   J0856.6-1105 & CGRABS J0856-1105 & UNKNOWN & Y\\
  136.222 & -57.584 & 0.98& 2.44 &  J090453-573504 &   J0905.1-5736 & PKS 0903-57 & FSRQ & N\\
  137.437 & -2.524 & 0.99 & 2.65 &  J090944-023129 &   J0909.6-0229 & PKS 0907-023 & FSRQ & Y\\
  148.261 & -8.671 & 0.99 & 1.58 &  J095302-084018 &   J0953.0-0838 & CRATES J0953-0840 & BLL & Y\\
  152.715 & -2.005 & 0.95 & 4.99 &  J101051-020019 &   J1011.0-0156 & CRATES J1010-0200 & FSRQ & Y\\
  156.640 & -85.720 & 0.89& 2.50 &  J102633-854315 &   J1028.7-8543 & PKS 1029-85 & UNKNOWN & Y\\
  164.679 & -80.064 & 0.98& 3.28 &  J105843-800353 &   J1058.1-8006 & PKS 1057-79 & BLL & Y\\
  164.801 & -11.572 & 0.97& 3.37 &  J105912-113422 &   J1059.3-1132 & PKS B1056-113 & BLL & Y\\
  165.968 & -53.950 & 0.99& 1.15 &  J110352-535700 &   J1103.9-5355 & PKS 1101-536 & UNKNOWN & N\\
  170.354 & -5.899 & 0.99 & 1.63 &  J112125-055356 &   J1121.5-0554 & PKS 1118-05 & FSRQ & Y\\
  170.831 & -64.293 & 0.97& 3.53 &  J112319-641735 &   J1122.9-6415 & PMN J1123-6417 & UNKNOWN & N\\
  171.381 & -35.950 & 0.97& 2.88 &  J112531-355703 &   J1125.5-3559 & CRATES J1125-3557 & UNKNOWN & Y\\
  171.768 & -18.955 & 0.97& 3.99 &  J112704-185719 &   J1126.8-1854 & PKS 1124-186 & FSRQ & Y\\
  172.529 & -14.824 & 0.98& 2.35 &  J113007-144927 &   J1130.2-1447 & PKS 1127-14 & FSRQ & N\\
  176.756 & -38.202 & 0.99& 0.65 &  J114701-381210 &   J1146.9-3812 & PKS 1144-379 & FSRQ & Y\\
  176.965 & -7.410 & 0.98 & 2.87 &  J114751-072438 &   J1147.7-0722 & PKS 1145-071 & FSRQ & Y\\
  178.072 & -8.684 & 0.93 & 4.26 &  J115217-084103 &   J1152.2-0836 & PKS B1149-084 & FSRQ & Y\\
  178.942 & -81.021 & 0.84& 9.22 &  J115546-810117 &   J1153.4-8108 & PMN-CA J1149-8112 & UNKNOWN & N\\
  178.525 & -32.712 & 0.98& 2.49 &  J115406-324243 &   J1154.2-3242 & PKS 1151-324 & UNKNOWN & Y\\
  179.794 & -21.834 & 0.96& 4.38 &  J115910-215005 &   J1159.4-2149 & PKS 1157-215 & FSRQ & N\\
  181.07 & -7.169 & 0.96  & 4.31 &  J120416-071009 &   J1204.3-0714 & CRATES J1204-0710 & BLL & Y\\
  181.389 & -26.568 & 0.93& 6.55 &  J120533-263404 &   J1205.9-2637 & PKS 1203-26 & FSRQ & Y\\
  191.695 & -25.796 & 0.98& 2.21 &  J124646-254749 &   J1246.7-2545 & PKS 1244-255 & FSRQ & Y\\
  194.046 & -5.789 & 0.99 & 0.43 &  J125611-054721 &   J1256.2-0547 & 3C 279 & FSRQ & Y\\
  194.066 & -11.776 & 0.94& 4.50 &  J125615-114635 &   J1256.5-1148 & CRATES J1256-1146 & UNKNOWN & Y\\
  194.659 & -18.000 & 0.95& 3.91 &  J125838-180001 &   J1258.4-1802 & PKS B1256-177 & FSRQ & Y\\
  194.727 & -22.325 & 0.96& 3.27 &  J125854-221930 &   J1258.7-2221 & PKS 1256-220 & FSRQ & Y\\
  195.917 & -46.350 & 0.92& 4.48 &  J130340-462103 &   J1304.0-4622 & CGRABS J1303-4621 & FSRQ & Y\\
  196.364 & -49.467 & 0.99& 0.24 &  J130527-492804 &   J1305.4-4928 & NGC 4945 & AGN$^{(b)}$ & Y\\
  197.071 & -67.117 & 0.86& 7.72 &  J130817-670704 &   J1307.3-6701 & PKS 1304-668 & UNKNOWN & N\\
  199.033 & -33.649 & 0.98& 2.60 &  J131608-333858 &   J1316.1-3341 & PKS 1313-333 & FSRQ & Y\\
  200.653 & -9.615 & 0.93 & 6.57 &  J132236-093655 &   J1322.7-0943 & OP -034 & FSRQ & Y\\
  201.365 & -43.018 & 0.98& 2.20 &  J132527-430104 &   J1325.6-4300 & CEN A & AGN & Y\\
  201.705 & -52.939 & 0.99& 2.38 &  J132649-525623 &   J1327.0-5257 & PMN J1326-5256 & UNKNOWN & N\\
  202.254 & -56.134 & 0.98& 3.32 &  J132901-560802 &   J1329.2-5605 & PMN J1329-5608 & UNKNOWN & N\\
  202.545 & -70.053 & 0.96& 4.29 &  J133010-700313 &   J1330.7-7006 & PKS 1326-697 & UNKNOWN & N\\
  203.018 & -5.162 & 0.96 & 3.53 &  J133204-050944 &   J1331.9-0506 & PKS 1329-049 & FSRQ & Y\\
  203.163 & -12.937 & 0.99& 0.32 &  J133239-125616 &   J1332.6-1255 & CRATES J1332-1256 & FSRQ & Y\\
  204.415 & -12.956 & 0.99& 1.98 &  J133739-125724 &   J1337.7-1255 & PKS 1335-127 & FSRQ & Y\\
  206.060 & -17.395 & 0.99& 0.63 &  J134414-172342 &   J1344.2-1723 & CGRABS J1344-1723 & FSRQ & Y\\
  206.919 & -37.843 & 0.97& 2.36 &  J134740-375037 &   J1347.8-3751 & CRATES J1347-3750 & FSRQ & Y\\
  208.694 & -10.684 & 0.99& 1.86 &  J135446-104103 &   J1354.9-1041 & PKS 1352-104 & FSRQ & Y\\
  210.174 & -56.082 & 0.93& 5.47 &  J140041-560455 &   J1400.9-5559 & PMN J1400-5605 & UNKNOWN & N\\
  211.915 & -43.042 & 0.95& 5.73 &  J140739-430231 &   J1407.5-4256 & CGRABS J1407-4302 & UNKNOWN & Y\\
  212.235 & -7.873 & 0.99 & 1.43 &  J140856-075225 &   J1408.9-0751 & PKS 1406-076 & FSRQ & Y\\
  216.984 & -42.105 & 0.97& 4.33 &  J142756-420618 &   J1428.2-4204 & PKS 1424-41 & FSRQ & Y\\
  220.988 & -39.144 & 0.99& 1.87 &  J144357-390839 &   J1444.0-3906 & PKS 1440-389 & BLL & Y\\
  224.361 & -35.653 & 0.99& 1.52 &  J145726-353910 &   J1457.5-3540 & PKS 1454-354 & FSRQ & Y\\
  226.258 & -34.545 & 0.97& 3.36 &  J150502-343242 &   J1505.1-3435 & CRATES J1505-3432 & UNKNOWN & Y\\
  227.723 & -5.718 & 0.96 & 4.69 &  J151053-054307 &   J1511.1-0545 & PKS 1508-05 & FSRQ & Y\\
  228.210 & -9.099 & 0.99 & 0.30 &  J151250-090558 &   J1512.8-0906 & PKS 1510-08 & FSRQ & Y\\
  228.667 & -47.808 & 0.93& 6.36 &  J151440-474828 &   J1514.1-4745 & PMN J1514-4748 & UNKNOWN & N\\
  229.424 & -24.372 & 0.99& 1.78 &  J151741-242220 &   J1517.8-2423 & AP LIB & BLL & Y\\
  230.657 & -27.503 & 0.98& 2.67 &  J152237-273011 &   J1522.6-2732 & PKS 1519-273 & BLL & Y\\
  238.381 & -24.368 & 0.98& 3.21 &  J155331-242206 &   J1553.4-2425 & PKS 1550-242 & UNKNOWN & Y\\
  238.389 & -31.308 & 0.99& 2.25 &  J155333-311832 &   J1553.5-3116 &	AT20G J1553-3118 & BLL & N\\
  240.961 & -49.068 & 0.99& 0.42 &   J160350-490405 &	J1603.8-4903 & PMN J1603-4904 & UNKNOWN & N\\
  241.129 & -44.692 & 0.98& 2.74 &  J160431-444131 &   J1604.7-4443 & PMN J1604-4441 & UNKNOWN & N\\
  242.693 & -66.816 & 0.99& 0.93 &  J161046-664900 &   J1610.6-6649 & CRATES J1610-6649 & BLL & Y\\
  242.591 & -39.982 & 0.92& 6.52 &  J161021-395857 &   J1610.8-3955 & VCS2 J1610-3958 & FSRQ & N\\
  244.324 & -58.801 & 0.95& 5.51 &  J161717-584806 &   J1617.7-5843 & PMN J1617-5848 & UNKNOWN & N\\
  244.455 & -77.288 & 0.99& 1.04 &  J161749-771718 &   J1617.9-7716 & PKS 1610-77 & FSRQ & Y\\
  246.445 & -25.460 & 0.98& 2.83 &  J162546-252739 &   J1625.7-2524 & PKS 1622-253 & FSRQ & Y\\
  246.525 & -29.857 & 0.95& 5.45 &  J162606-295126 &   J1626.2-2956 & PKS 1622-29 & FSRQ & Y\\
  247.227 & -61.876 & 0.93& 6.45 &  J162854-615236 &   J1629.5-6147 & PMN J1628-6152 & UNKNOWN & N\\
  252.569 & -50.746 & 0.99& 2.78 &  J165016-504446 &   J1650.4-5042 & PMN J1650-5044 & UNKNOWN & N\\
  255.901 & -62.210 & 0.86& 7.63 &  J170336-621238 &   J1702.7-6217 & CGRABS J1703-6212 & FSRQ & N\\
  259.400 & -33.701 & 0.93& 5.08 &  J171736-334206 &   J1717.9-3343 & TXS 1714-336 & BLL & N\\
  270.677 & -39.668 & 0.98& 1.97 &  J180242-394007 &   J1802.5-3939 & BZU J1802-3940 & UNKNOWN & N\\
  278.129 & -56.989 & 0.97& 1.34 &  J183231-565920 &   J1832.6-5700 & CRATES J1832-5659 & BLL & Y\\
  278.416 & -21.061 & 0.99& 0.60 &  J183339-210341 &   J1833.6-2103 & PKS 1830-21 & FSRQ & N\\
  282.357 & -43.236 & 0.99& 1.99 &  J184925-431412 &   J1849.6-4314 & CRATES J1849-4314 & BLL & Y\\
  287.790 & -20.115 & 0.99& 1.11 &  J191109-200655 &   J1911.2-2007 & PKS B1908-201 & FSRQ & Y\\
  289.437 & -19.358 & 0.99& 0.47 &  J191744-192131 &   J1917.7-1922 & CGRABS J1917-1921 & BLL & Y\\
  289.566 & -41.192 & 0.99& 3.96 &  J191816-411131 &   J1918.4-4108 & CRATES J1918-4111 & BLL & Y\\
  290.35 & -12.531 & 0.96 & 4.86 &  J192124-123154 &   J1921.1-1234 & CRATES J1921-1231 & UNKNOWN & Y\\
  290.884 & -21.075 & 0.99& 0.07 &  J192332-210433 &   J1923.5-2104 & OV -235 & FSRQ & Y\\
  291.263 & -10.303 & 0.99& 1.14 &  J192503-101812 &   J1925.1-1018 & CRATES J1925-1018 & BLL & Y\\
  294.317 & -39.967 & 0.80& 11.4 &  J193716-395801 &   J1938.2-3957 & PKS 1933-400 & FSRQ & N\\
  296.497 & -31.193 & 0.90& 7.22 &  J194559-311138 &   J1946.1-3118 & PKS 1942-313 & BLL & Y\\
  298.671 & -11.389 & 0.98& 2.55 &  J195441-112323 &   J1954.8-1124 & CGRABS J1954-1123 & FSRQ & Y\\
  299.805 & -42.768 & 0.97& 4.39 &  J195913-424607 &   J1959.3-4241 & CGRABS J1959-4246 & FSRQ & Y\\
  300.237 & -17.815 & 0.99& 0.90 &  J200057-174857 &   J2000.9-1749 & PKS 1958-179 & FSRQ & Y\\
  301.98 & -44.578 & 0.96 & 4.38 &  J200755-443444 &   J2007.9-4430 & PKS 2004-447 & AGN$^{(d)}$ & Y\\
  302.102 & -4.308 & 0.96 & 3.51 &  J200824-041829 &   J2008.6-0419 & 3C 407 & AGN & Y\\
  302.356 & -48.831 & 0.99& 0.80 &  J200925-484953 &   J2009.5-4849 & PKS 2005-489 & BLL & Y\\
  303.619 & -0.789 & 0.99 & 1.27 &  J201428-004723 &   J2014.5-0047 &	AT20G J2014-0047 & BLL & N\\
  303.813 & -1.625 & 0.80 & 7.99 &  J201515-013731 &   J2015.3-0129 & PKS 2012-017 & BLL & Y\\
  306.419 & -7.597 & 0.99 & 0.13 &  J202540-073551 &   J2025.6-0735 & PKS 2023-07 & FSRQ & Y\\
  306.473 & -28.763 & 0.86& 7.04 &  J202553-284547 &   J2025.9-2852 & CGRABS J2025-2845 & UNKNOWN & Y\\
  314.068 & -47.246 & 0.99& 0.84 &  J205616-471447 &   J2056.3-4714 & PKS 2052-47 & FSRQ & Y\\
  315.909 & -62.540 & 0.94& 5.46 &  J210338-623225 &   J2103.9-6237 & CRATES J2103-6232 & UNKNOWN & Y\\
  321.627 & -46.096 & 0.95& 4.56 &  J212630-460548 &   J2126.1-4603 & PKS 2123-463 & FSRQ & Y\\
  323.542 & -1.887 & 0.86 & 10.3 &  J213410-015316 &   J2134.0-0203 & 4C -02.81 & FSRQ & Y\\
  323.833 & -50.114 & 0.88& 10.8 &  J213520-500651 &   J2135.8-4957 & CRATES J2135-5006 & FSRQ & Y\\
  324.850 & -42.588 & 0.99& 0.08 &  J213924-423520 &   J2139.3-4235 & CRATES J2139-4235 & BLL & Y\\
  326.251 & -33.954 & 0.96& 5.07 &  J214501-335716 &   J2145.4-3358 & CGRABS J2145-3357 & FSRQ & Y\\
  329.526 & -15.019 & 0.96& 3.02 &  J215806-150109 &   J2157.9-1503 & PKS 2155-152 & FSRQ & Y\\
  329.717 & -30.225 & 0.99& 0.55 &  J215852-301332 &   J2158.8-3013 & PKS 2155-304 & BLL & Y\\
  331.680 & -0.517 & 0.86 & 11.6 &  J220643-003103 &   J2207.1-0021 & CGRABS J2206-0031 & BLL & N\\
  331.932 & -53.776 & 0.97& 2.55 &  J220743-534633 &   J2207.8-5344 & PKS 2204-54 & FSRQ & Y\\
  333.260 & -25.491 & 0.98& 1.35 &  J221302-252930 &   J2213.1-2529 & PKS 2210-25 & FSRQ & Y\\
  336.447 & -4.950 & 0.99 & 1.55 &  J222547-045701 &   J2225.8-0457 & 3C 446 & FSRQ & Y\\
  337.416 & -8.548 & 0.99 & 1.87 &  J222940-083254 &   J2229.7-0832 & PKS 2227-08 & FSRQ & Y\\
  339.141 & -14.556 & 0.99& 1.44 &  J223634-143322 &   J2236.4-1432 & PKS 2233-148 & BLL & Y\\
  339.283 & -39.360 & 0.96& 2.50 &  J223708-392137 &   J2237.2-3919 & CRATES J2237-3921 & FSRQ & Y\\
  340.860 & -25.742 & 0.93& 5.20 &  J224326-254431 &   J2243.1-2541 & PKS 2240-260 & BLL & Y\\
  342.685 & -28.111 & 0.97& 3.56 &  J225044-280639 &   J2250.8-2809 & CGRABS J2250-2806 & AGN & Y\\
  344.171 & -20.194 & 0.97& 4.96 &  J225641-201141 &   J2256.3-2009 & PKS 2254-204 & BLL & Y\\
  344.524 & -27.972 & 0.96& 1.63 &  J225805-275821 &   J2258.0-2757 & PKS 2255-282 & FSRQ & N\\
  348.935 & -50.310 & 0.92& 4.82 &  J231544-501839 &   J2315.9-5014 & PKS 2312-505 & BLL & Y\\
  350.883 & -3.284 & 0.98 & 1.72 &  J232331-031705 &   J2323.5-0315 & PKS 2320-035 & FSRQ & Y\\
  351.362 & -48.004 & 0.97& 2.56 &  J232526-480017 &   J2325.6-4758 & PKS 2322-482 & BLL & Y\\
  352.337 & -49.927 & 0.99& 1.67 &  J232921-495539 &   J2329.2-4954 & PKS 2326-502 & FSRQ & Y\\
  352.766 & -21.804 & 0.98& 2.97 &  J233104-214815 &   J2331.0-2145 & CRATES J2331-2148 & FSRQ & Y\\
  354.489 & -2.515 & 0.87 & 6.36 &  J233757-023057 &   J2338.3-0231 & PKS 2335-027 & FSRQ & Y\\
  357.010 & -16.519 & 0.99& 1.88 &  J234802-163111 &   J2348.0-1629 & PKS 2345-16 & FSRQ & Y\\
\hline
%\end{tabular}
%\vskip 0.4 true cm
\caption{222 associations found from the cross correlation of the 1FGL 
sample and the AT20G survey and classified as blazars in the 1LAC sample. 
The position of the AT20G radio counterpart, the probability of the association, 
the angular separation between the radio counterpart and the \fe\ 
1FGL centroid are given. The classification of the source in the 
1LAC (FSRQ, BLL or UNKNOWN) is given with the name of the counterpart
 associated in the 1LAC sample. The last column is a flag: Y=the source 
is classified as a high confidence blazar in the 1LAC sample and it 
belongs to the ``clean" sub--sample (see A10a for details). \\
$^{(a)}$ NGC 253, starburst galaxy, (Abdo et al., 2010b). \\
$^{(b)}$ NGC 4945, starburst/Seyfert 2 (Lenc  \& Tingay, 2009). \\ 
$^{(c)}$ PKS 0625--35, Low--Excitation FRI radio galaxy (Gliozzi  et al., 2008). \\
$^{(d)}$ PKS 2004--447, narrow--Line Seyfert 1 (Abdo et al., 2009b). 
}
\label{tab3}
\end{longtable}
\end{center}
% ----------------------------------------------------------------------------------------------------------------------------------

\end{document}